\documentstyle[12pt,epsfig]{article}
\begin{document}
\thispagestyle{empty}
\newcommand{\be}{\begin{equation}}
\newcommand{\ee}{\end{equation}}
\newcommand{\ba}{\begin{eqnarray}}
\newcommand{\ea}{\end{eqnarray}}
\date{}
\title{\mbox{The Fokker-Planck equation for bistable potential}
in the optimized expansion}
\author{Anna Okopi\'nska\\
 Institute of Physics, Pedagogical University,\\
 Konopnickiej 15, 25-406 Kielce, Poland\\
 and \\Institute f\"{u}r Theoretische Physik,\\ Freie Universit\"{a}t
 Berlin,\\
 Arnimallee 14, D-14195 Berlin, Germany\\
 }
\maketitle
\begin{abstract}
The optimized expansion is used to formulate a systematic
approximation scheme to the probability distribution of a
stochastic system. The first order approximation for the
one-dimensional system driven by noise in an anharmonic potential is
shown to agree well with the
exact solution of the Fokker-Planck equation. Even for a bistable
system the whole
period of evolution to equilibrium is correctly described at
various noise intensities.
\end{abstract}

\section{Introduction}
The Fokker-Planck (FP) equation is widely used to describe
non-equilibrium systems in physics, chemistry and
biology~\cite{FP}. The stochastic approach consists in
representing the most relevant degrees of freedom of the system by
the variable $x$ driven by noise and deterministic interaction
potential $U(x,t)$. The time development of the probability
distribution $W(x,t)$ is given by a partial differential equation
\begin{equation}\label{FP}
{\frac{\partial}{\partial t }}W(x,t)=\mathbf{L_{FP}}W(x,t)
:={\frac{\partial}{\partial x }}[U'(x,t)W(x,t)+D
{\frac{\partial^{2}}{\partial x^{2} }}W(x,t)
\end{equation}
where the diffusion coefficient, $D$, represents a noise intensity
and $U'(x,t)$, denoting the derivative of the interaction
potential with respect to $x$, is called the drift coefficient.
The Green's function of the FP equation, $P(x,t|x',t')$ which
fulfils the initial condition $P(x,t|x',t')=\delta(x,x')$, is
called the transition probability (conditional
probability), since it describes the evolution of the probability
density from time $t'$ to $t$:
\begin{equation}\label{PK}
W(x,t)= \int P(x,t|x',t')W(x',t')dx'.
\end{equation}
For a time independent potential, $U(x)$, the separation ansatz
\begin{equation}\label{tdep}
  W(x,t)=\Phi(x) e^{-\kappa t}
\end{equation}
reduces the time dependent FP equation(~\ref{FP}) to the
stationary eigenfunction equation
\begin{equation}\label{FPst}
\mathbf{L_{FP}}\Phi(x)={\frac{\partial U'(x)\Phi(x)}{\partial x
}}+D {\frac{\partial^{2}\Phi(x)}{\partial x^{2} }}= -\kappa
\Phi(x).
\end{equation}
The lowest eigenvalue of a FP operator is identically zero,
$\kappa=0$, and the corresponding eigenfunction $\Phi_{0}(x)$ can
be found exactly, yielding the stationary probability distribution
of the form
\begin{equation}\label{Wst}
W_{st}(x)=\Phi_{0}(x)=N e^{-\frac{U(x)}{D}}
\end{equation}
with the normalization constant
$N=({\int_{-\infty}^{\infty}e^{-\frac{U(x)}{D}}dx})^{-1}$. For an
arbitrary potential $U(x)$ the higher eigenfunctions and the
non-stationary probability distribution cannot be found exactly,
there is thus a need to develop approximation methods. To this end
it is convenient to transform the FP equation into the
Schr\"odinger equation~\cite{FP}. The transformation
\begin{equation}
\Psi(x)=e^{\frac{U(x)}{2D}}\Phi(x)
\end{equation}
brings the FP operator to the Hermitian form
\begin{equation}\label{Her}
{\mathcal{L}}=e^{\frac{U(x)}{2D}}L_{FP}e^{\frac{-U(x)}{2D}}=
  D\frac{d^{2}}{dx^{2}}-V(x)
\end{equation}
where
\begin{equation}\label{VS}
V(x)=\frac{[U'(x)]^{2}}{4D}-\frac{U''(x)}{2}.
\end{equation}
The operator ${\hbar
\mathcal{L}}$ has the
same form as the negative
Hamiltonian operator for the quantum mechanical particle of the
mass $M=\frac{\hbar}{2D}$ in the potential $\hbar V(x)$. The
transformed FP equation
\begin{equation}\label{psS}
  {\mathcal{L}}\Psi(x)=\left(D\frac{d^{2}}{dx^{2}}-V(x)\right)\Psi(x)=
  \lambda\Psi(x)
\end{equation}
is called a pseudo-Schr\"{o}dinger equation, because by
(\ref{tdep}) the wave function, $\Psi$, evolves in imaginary time
$t_{S}=-i\hbar t$. The transition probability, $P(x,t|x',t')$,
being the Green's function of the original FP equation~(\ref{FP}),
is related to the imaginary time evolution amplitude,
$K(x,t;x',t')$, of the pseudo-Schr\"{o}dinger equation~(\ref{psS})
by
\begin{equation}\label{PKao}
P(x,t|x',t')=\exp\left(\frac{U(x')-U(x)}{2D}\right)K(x,t;x',t').
\end{equation}
The evolution amplitude can be represented as a path integral
\be\label{KK} K(x,t|x',t')\!=\!
\!\int_{x(t')\!=\!x'}^{x(t)\!=\!x}\!\!Dxe^{\!-\!
\int_{t'}^{t}\!L[x]\!dt} \label{K} \ee over all functions which
begin at $x(t')=x'$ and end at $x(t)=x$, where the
pseudo-Schr\"{o}dinger Lagrangian of a particle is given by
\begin{equation}
L[x]\!=\!\frac{\dot x^2}{4D}\!+\!V(x).
\end{equation}
We shall study a stochastic system in an anharmonic potential, \be
U(x)\!=\!\frac{\gamma x^2}{2}\!+\!\lambda x^4. \ee In this case
the pseudo-Schr\"{o}dinger potential~(\ref{VS}) takes the form:
\be\label{VSao}
V(x)=-\frac{\gamma}{2}+\left(\frac{\gamma^2}{4D}-6\lambda\right)
x^2+\frac{2\lambda \gamma}{D} x^4+4\frac{\lambda^2x^6}{D}.\ee For
vanishing $\lambda$ the problem reduces to the exactly solvable
Ornstein-Uhlen\-beck process in a quadratic interaction potential,
$U(x)=\frac{\gamma x^2}{2}$. In this case the
pseudo-Schr\"{o}dinger potential is also quadratic,
$V(x)=-\frac{\gamma}{2}+\frac{\gamma^2}{4D}x^2$, and the path
integral for the evolution amplitude~(\ref{KK}) can be performed
exactly yielding \ba\label{KHO}
&&K_{\gamma}(x,t|x',t')=\sqrt{\frac{\gamma}{2\pi D(1-e^{-2\gamma
(t-t')})}}\nonumber\\&&\times
\exp{\left\{\frac{\gamma}{4D\sinh(\gamma (t-t'))}\left[(x^2+x'^2)
\cosh(\gamma (t-t'))-2xx'\right]\right\}}.\ea By~(\ref{PK}) this
leads to an exact expression for the transition probability of the
Ornstein-Uhlenbeck process \be\label{PHO}
P_{\gamma}(x,t|x',t')={\sqrt\frac{\gamma}{2\pi D(1-e^{-2\gamma
(t-t')})}} \exp \left(- \frac{\gamma(x-x'e^{-\gamma
(t-t')})^2}{2D(1-e^{-2\gamma (t-t')})}\right).\ee In the presence
of anharmonicity ($\lambda\ne 0$) the evolution amplitude,
$K(x,t;x',t')$, cannot be obtained exactly, but various
approximation methods are developed. The perturbative calculations
of $P(x,t|x',t')$ in powers of $\lambda$ are possible if
$\gamma>0$. In this case the first order approximation describes
well the evolution of the system approaching the stationary
distribution in the long time limit, only the normalization
worsens with increasing $\lambda$. However, for $\gamma<0$ the
method becomes inapplicable, since the transition probability is
non-normalizable. In this case the perturbative approximations
give a wrong description of the time evolution, since the maxima
of the transition probability escape to $x=\pm\infty$.

Few years ago we proposed the optimized expansion (OE) scheme for
the evolution amplitude~\cite{AOP} which has much better
convergence properties than the perturbative expansion in powers
of $\lambda$. Here we apply this method to generate a systematic
approximation scheme for the transition amplitude $P(x,t|x',t')$
by~(\ref{PK}). We will show that the method can be successfully
applied for describing the time evolution of a stochastic systems.
The efficiency of the method will be shown on the example of
bistable system driven by noise in the double-well potential. Such
a system attracts much attention in nonlinear optics (statistical
properties of laser light above threshold), solid-state physics
and chemistry. The solution of the FP equation which describes the
evolution from an intrinsically unstable state to the final
stationary state is of a special interest, and various
approximation methods have been proposed and compared with the
numerical solution in this case. However, in all the methods the
evolution process is divided into few steps and different
approximations are used in each time sector~\cite{Hu}. We will
show that the first order result of the OE provides a simple
approximation which agrees well with the exact solution of the FP
equation in the whole period of evolution to equilibrium at
various noise intensities.

\section{The optimized expansion}
\label{oe} The optimized expansion (OE) has been formulated as a
method to generate non-perturbative approximations for the
effective action in quantum field theory~\cite{AO}. The method
consists in calculating the effective action as a series in
$\epsilon$, by splitting the Lagrangian into \be
L=L_{0}+\epsilon(L- L_{0}), \ee where the unperturbed part
contains arbitrary parameters, which are optimized in every order
calculation (upon setting $\epsilon=1$). The method is equivalent
to a systematic re-summation of the perturbation series and gives
the Hartree-Fock-Bogolubov approximation in the leading order. A
similar idea has been applied to formulate systematic
approximation methods for other physical quantities in a number of
works ~\cite{OE,book} under different names (self-similar
perturbation theory, delta expansion, variational perturbation
theory, optimized perturbation theory...). The approach provides a
method to systematically improve any self-consistent approximation
of the theory. The OE has been also applied to the quantum
mechanics of the particle in the potential $V(x)$ by modifying the
classical Lagrangian to the form
\begin{equation}
L[x]\!=L_{\omega}+\epsilon V_{int}=\!\frac{M\dot x^2}{2}\!+
\frac{M\omega^2 x^2}{2}+ \epsilon\left(V(x)-\frac{M\omega^2
x^2}{2}\right),
\end{equation}
where the harmonic oscillator of the mass $M$ and an arbitrary
frequency $\omega$ is chosen as the unperturbed system, and
calculating physical quantities as a series in powers of
$\epsilon$. It has been shown, that for the ground state energy of
the quantum mechanical anharmonic oscillator a convergent series
is obtained~\cite{conv,book}, as opposed to perturbative series
which is asymptotic. Applying the method to calculate the
imaginary time evolution amplitude~\cite{AOP,vp}, it is convenient
to represent the series for $W(x,t;x',t')=\ln K(x,t;x't')$ by the
cumulant expansion. For the particle of the mass
$\frac{\hbar}{2D}$ we have
\begin{eqnarray}
W(x,t;x',t')&\!=\!&W_{\omega}(x,t;x',t')\!-\! \epsilon
<V_{int}(x)>_{\omega}\nonumber\\&\!+\!&\frac{\epsilon^2}{2}
(<V_{int}^2(x)>_{\omega} -
<V_{int}(x)>_{\omega}<V_{int}(x)>_{\omega})-... \label{Wn}
\end{eqnarray}
where by ~(\ref{KHO})\ba &&W_{\omega}(x,x',t,t')=\ln
K_{\omega}(x,x',t,t')=\nonumber\\&&\frac{1}{2}\ln
\left(\frac{\omega} {4\pi D\sinh\omega (t-t')}\right) -
\frac{\omega\left[(x^2+x'^2)\cosh\omega (t-t')-2
xx'\right]}{4D\sinh \omega (t-t')} \label{W0} \ea and the
expectation values are calculated for the unperturbed Lagrangian
\be <...>_{\omega}=\!
\!\int_{x(t')\!=\!x'}^{x(t)\!=\!x}\!\!Dx...e^{\!-\!
\int_{t'}^{t}\!\!L_{\omega}[x]\!dt}.\label{ev} \ee The $N$-th
order approximant, $W^{(N)}(x't';x,t)$, is obtained by truncating
the series~(\ref{Wn}) after the $N$-th term and setting
$\epsilon=1$, since only in this case does the modified action
agree with the classical one. The exact result, being a sum of an
infinite series, would not depend on arbitrary frequency, but any
finite order truncation shows such a dependence. We choose,
therefore, the value of the unperturbed frequency, $\omega$, to
make the given order approximant as insensitive as possible to its
small variation, by requiring \be \frac{\delta W^{(N)}}{\delta
\omega}=0.\label{gap} \ee The optimization condition determines
the value of $\omega$ as a function of $\beta, x$ and $x'$, which
changes from order to order, improving the convergence properties
of the approximation scheme.

In the case of polynomial potential the expectation values in
$W^{(N)}$~(\ref{Wn}) are given by Gaussian functional integrals
which can be easily performed, yielding an analytic expression for
$W^{(N)}$. The first order result for the evolution amplitude of
the quartic oscillator has been obtained in Ref.~\cite{AOP}, and
shown to give a satisfactory approximation to the particle density
in the broad range of the oscillator parameters. Here we have to
calculate the evolution amplitude in the potential
\begin{equation}
V(x)=g_{0}+g_{2} x^2+g_{4} x^4+g_{6} x^6,
\end{equation}
because the pseudo-Schr\"{o}dinger potential corresponding to the
bistable FP equation~(\ref{VSao}) is sextic. After setting
$\epsilon=1$ the first order results can be written as \ba
\label{W6}&&W^{(1)}(x,t;x',t'))\!=\!-t g_{0}+
W_{\omega}(x,t;x',t')\!+\! \left(
\frac{\omega^2}{2}\!-\!g_{2}\right)
\int_{t'}^{t}[L^2(\tau)+K(\tau)]d\tau
\nonumber\\
&&-g_{4}\int_{t'}^{t} [L^4(\tau)\!+\!6
L^2(\tau)K(\tau)\!+\!3K^2(\tau)]d\tau\! \nonumber\\&&-
g_{6}\int_{t'}^{t}
[L^6(\tau)\!+\!15K(\tau)L^4(\tau)\!+\!45K^2(\tau) L(\tau)^2\!+\!
15K^3(\tau)]d\tau\label{Wi} \ea where \[
L(\tau)\!=\!\frac{x\sinh\omega (\tau\!-\!t')\!+\!x_{b}\sinh\omega
(t\!-\!\tau)}{\sinh\omega (t\!-\!t')}
{\mbox{\rm~and~}}K(\tau)\!=\!\frac{2D\sinh \omega
(\tau\!-\!t')\sinh\omega (t\!-\!\tau)}{\omega\sinh \omega
(t\!-\!t')} \] and the optimization condition~(\ref{gap}) reduces
to
\begin{eqnarray}\label{gap6}
\left(\!g_{2}\!-\!\frac{\omega^2}{2}\!\right)
\!\frac{\partial}{\partial\omega^2}\!\int_{t'}
^{t}[L^2(\tau)\!+\!K(\tau)]d\tau\!+\!g_{4}\frac{\partial}
{\partial\omega^2}\!\int_{t'}^{t} [L^4(\tau)\!+\!6 L^2(\tau)
K(\tau)\!+\!3
K^2(\tau)]d\tau\nonumber\\+g_{6}\frac{\partial}{\partial\omega^2}
\int_{t'}^{t}[L^6(\tau)+ 15K(\tau)L^4(\tau) + 45K^2(\tau)
L^2(\tau) + 15K^3(\tau)]d\tau=0.~~~~~~~
\end{eqnarray}
Upon performing the integrals in the above expressions an analytic
expression for the evolution amplitude
$K(x,t|x',t')=e^{W(x,t;x',t')}$ is obtained. The result is
optimized by choosing the value of the frequency to
fulfil~(\ref{gap6}) and used to calculate the first order
approximation to the transition probability $P(x,t|x',t')$
according to Eq.\ref{PK}.

After completing this work we learned that approximate solutions
of the FP equation in an anharmonic potential have been also
studied by directly improving the perturbative expansion of the
transition probability, using a drift coefficient as a variational
parameter~\cite{KPP}. The variational expressions substantially
differ from ours, because the sextic term of the
pseudo-Schr\"{o}dinger potential ~(\ref{VSao}) does not contribute
to the first order in their method. The numerical results are also
different, especially in the case of the bistable potential.

\section{Results and conclusions}
\label{con}

The approximation obtained in the first order of the OE is exact
at $\lambda=0$, and it gives very good results when the anharmonic
terms are small and the pseudo-Schr\"{o}dinger potential is convex
($\gamma>0$). The shape of the transition probability is well
described, but the total probability slowly decreases in time. A
similar problem appears in the first order calculation in the
variational perturbation method ~\cite{KPP}, where the total
probability increases in time. The spoiling of the normalization
by optimization is a general feature of perturbative schemes with
variational parameters~\cite{Kuni} and can be cured by normalizing
the each order result by hand.
The normalized
probability distribution obtained in the first order of the OE shows
a good agreement with the exact solution of the FP equation in the whole
period of evolution. Here we show the results in the most demanding case
of double-well interaction potential ($\gamma<0$), when the
pseudo-Schr\"{o}dinger potential has a multiple-well structure.
The critical value of the diffusion coefficient
($D_{cr}=\frac{\gamma^2}{24\lambda})$, distinguishes two cases:
the pseudo-Schr\"{o}dinger potential of double-well shape in the
case of large noise ($D>D_{cr}$), and of triple-well shape in the
case of small noise.

We compare the results of our approximation to the transition
probability, $P(x,t|0,0)$, with the exact results calculated
numerically in Ref.~\cite{Hu} and Ref.~\cite{Liu} for the
interaction potential $U(x)=-\frac{1}{2}x^2+\frac{1}{4}x^4$, which
corresponds to $\gamma=-1$ and $\lambda=\frac{1}{4}$ in our
notation. The critical value of the diffusion coefficient is
$D_{cr}=\frac{1}{6}$ in this case. The results for the time
evolution of the transition probability at noise intensity $D=0.1$
are shown in Fig.1, and the results at $D=0.05$ and $D=0.01$, in
Fig.2 and Fig.3, respectively. The pseudo-Schr\"{o}dinger
potential for the considered values of $D$ is also plotted.

The transition probability for the system being initially in the
unstable state ($P(x,0|0,0)=\delta (x-0)$) is presented at
different times of the evolution: the first value is in the
initial time region, the next ones are in the intermediary region,
and the last one is in the final time region when the stationary
distribution is already achieved. It is remarkable that a good
description of the evolution from the unstable configuration to
the stable one is obtained already in the first order of the OE,
even in the difficult case of small noise. This is due to the
optimization of the variational parameter $\omega$ at a given
point $x$ in the considered time $t$. One has to stress, that in
our calculation the stationarity condition, which determines the
optimal value of $\omega$ (\ref{gap6}), has a solution for all the
values of $x$ during the whole period of evolution. This is in
difference with the variational perturbation method~\cite{KPP},
where different criteria of the variational parameter fixing are
used in the initial and in the final regions, and the worst
description is obtained in the intermediate stage of evolution.
In our
approximation the initial and intermediate stage of evolution are
described very
well for all values of noise, but the discrepancies appear in the
final stage of evolution and the asymptotic distribution differs
from the exact one. For $D=0.05$ and $D=0.01$ the approximation is
of similar quality as the two or three-stage approximations based
on $\Omega$ expansion~\cite{Hu}, and gives also a good description
for $D=0.1$, when the approximation discussed by Hu becomes
inaccurate as observed in Ref.~\cite{Liu}. The accuracy of our
approximation can be improved by higher order calculation in a
systematic way.

One has to note that an extension of this approach to higher
dimensional systems is
possible. Also the dynamics of a bistable stochastic system driven
by time-dependent forces can be studied in the OE. The influence
of a periodic force on the bistable system, which is a topic of
current interest because of the phenomenon of stochastic
resonance, will be discussed in a future publication.

\section{Acknowledgment}
The author thanks to Axel Pelster and Hagen Kleinert for useful
discussions and showing her the results of their work before
publication. The support from DAAD is gratefully acknowledged.

\newpage
\begin{figure}
\setlength{\unitlength}{1cm}
\centerline{\epsfxsize=16cm\epsfbox{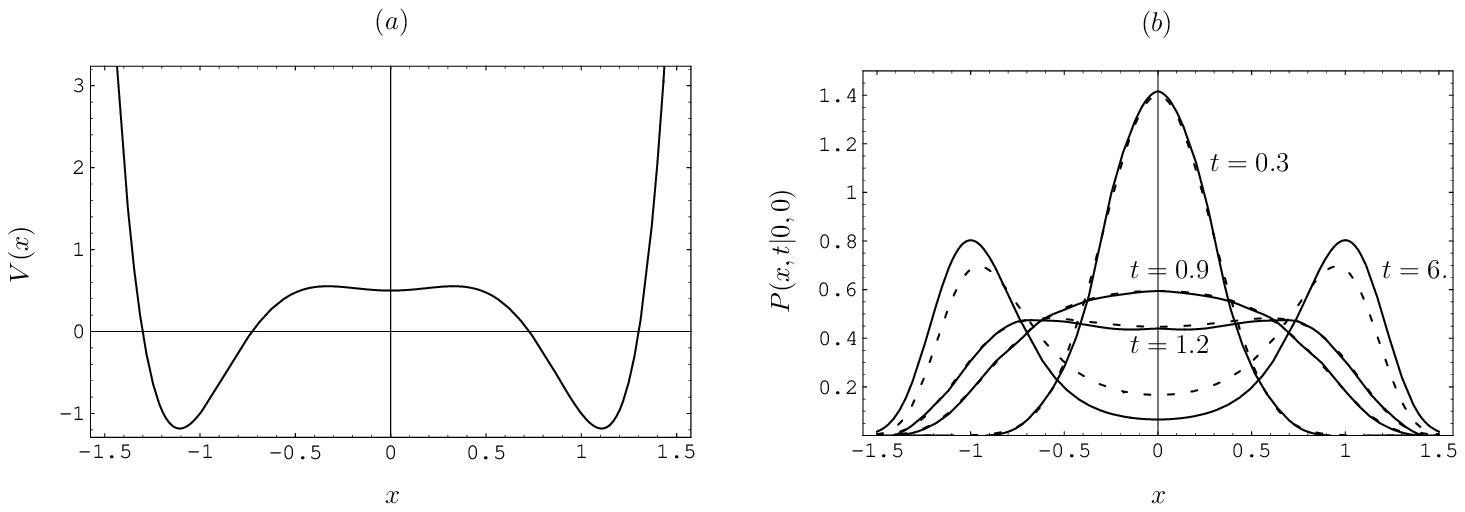}}
\caption{The
pseudo-Schr\"{o}dinger potential {\textit(a)} and the
time evolution of $P(x,t|0,0)$ {\textit(b)} for the bistable
potential \mbox{$U(x)=-\frac{x^2}{2}+x^4$}
at $D=0.1$ in the
first order of the OE ({\it dashed line})
at $t=0.3$, $t=0.9$, $t=1.2$ and $t=6.$,
compared with the exact results taken from
Ref.~\cite{Liu} ({\textit solid line})
}
\end{figure}
\begin{figure}
\setlength{\unitlength}{1cm}
\centerline{\epsfxsize=16cm\epsfbox{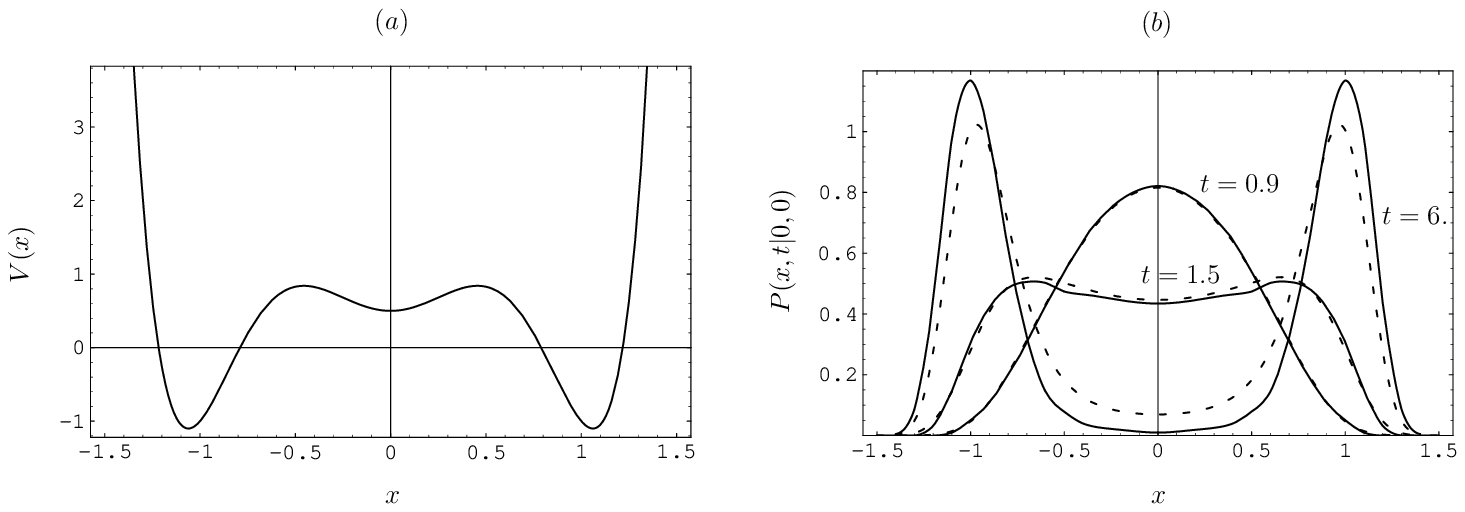}}
\caption{Same as in Fig.1, but for $D=0.05$. at the time $t=0.9$,
$t=1.5$ and $t=6.$, compared with the exact results taken from
Ref.~\cite{Hu} (\textit{ solid line}).}
\end{figure}
\begin{figure}

\setlength{\unitlength}{1cm}
\centerline{\epsfxsize=16cm\epsfbox{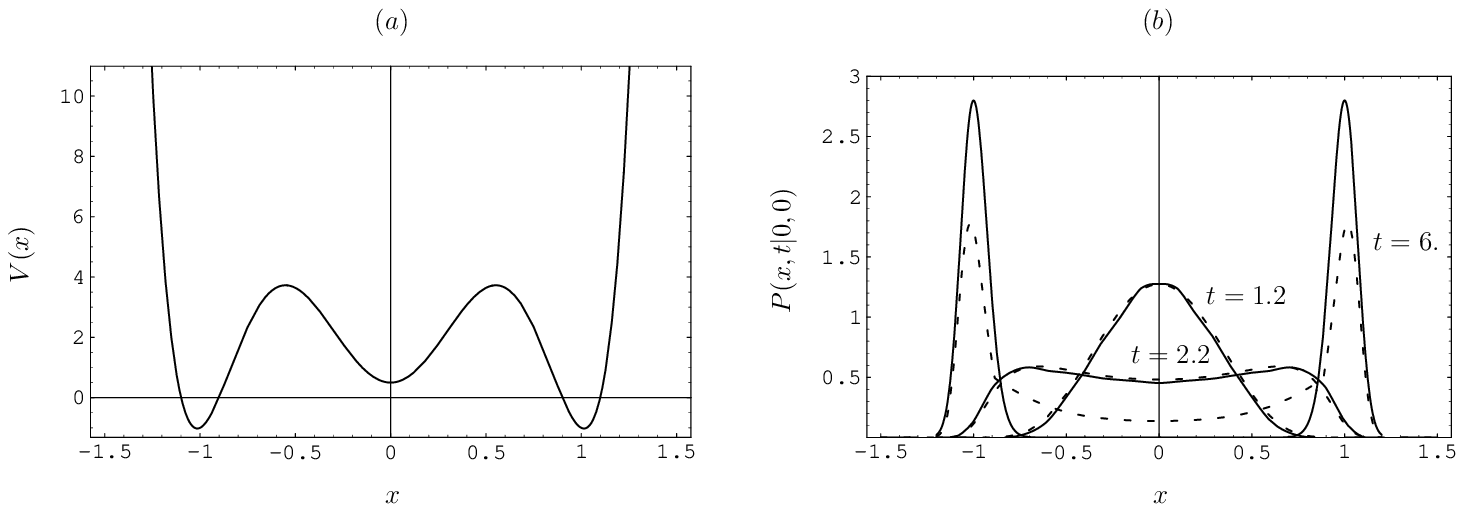}}

\caption{Same as in Fig.1, but for $D=0.01$. at the time $t=1.2$,
$t=2.2$ and $t=6.1$, compared with the exact results taken from
Ref.~\cite{Hu} {\it solid line}).}

\end{figure}

\end{document}